\documentclass[journal,twoside,web]{ieeecolor}
\usepackage{tmi}
\usepackage{cite}
\usepackage{amsmath,amssymb,amsfonts}

\usepackage{graphicx}
\usepackage{textcomp}
\usepackage{orcidlink} 
\usepackage{times}
\usepackage{epsfig}
\usepackage{graphicx}
\usepackage{amsmath}
\usepackage{amssymb}
\usepackage[ruled,vlined,linesnumbered]{algorithm2e}
\usepackage{booktabs}
\usepackage{caption}
\usepackage{subcaption}
\usepackage{pgfplots}
\pgfplotsset{compat=1.16}
\usepackage{comment}
\usepackage{multirow}
\usepackage{array, makecell} 
\usepackage{xcolor}
\usepackage{lettrine}
\usepackage{physics} 
\usepackage{acronym}
\acrodef{LBP}[LBP]{Low Back Pain}
\acrodef{2D}[2D]{Two-Dimensional}
\acrodef{FCN}[FCN]{Fully Convolutional Network}
\acrodef{GAME}[GAME]{Grid Average Mean Absolute Error}
\acrodef{DL}[DL]{Deep Learning}
\acrodef{US}[US]{Ultrasound}
\acrodef{CT}[CT]{Computerized Tomography}
\acrodef{TrA}[TrA]{Transversus Abdominis}
\acrodef{ICC}[ICC]{Intraclass Correlation Coefficient}
\acrodef{CNN}[CNN]{Convolutional Neural Network}
\acrodef{JCU}[JCU]{James Cook University}
\acrodef{ADIM}[ADIM]{Abdominal Drawing-In Maneuver}
\acrodef{MWA}[MWA]{Muscle Width Agreement}
\acrodef{MAE}[MAE]{Mean Absolute Error}
\acrodef{LCFCN}[LCFCN]{Localization-based Counting Fully Convolutional Network}
\usepackage[switch]{lineno}
\usepackage{everypage}
\usepackage{tikz}
\usepackage{lipsum}

\usepackage{moreverb,url}
\usepackage[linesnumbered,ruled,vlined]{algorithm2e}

\usepackage{algorithmic}

\usepackage[colorlinks,
            linkcolor=blue,
            anchorcolor=blue,
            citecolor=blue]{hyperref}

\begin{document}


\title {A Two-Stage Generative Model with CycleGAN and Joint Diffusion for MRI-based Brain Tumor Detection}
\author{Wenxin Wang\textsuperscript{\orcidlink{0000-0002-5927-7456}}, Zhuo-Xu Cui\textsuperscript{\orcidlink{0000-0001-9283-881X}},Guanxun Cheng, Chentao Cao\textsuperscript{\orcidlink{0000-0002-3974-3413}}, Xi Xu\textsuperscript{\orcidlink{0000-0001-8726-5834}}, Ziwei Liu, Haifeng Wang \textsuperscript{\orcidlink{0000-0003-4229-3668}}, \IEEEmembership{Member, IEEE}, Yulong Qi, Dong Liang\textsuperscript{\orcidlink{0000-0001-6257-0875}} and Yanjie Zhu\textsuperscript{\orcidlink{0000-0002-9131-689X}}
\thanks{ Manuscript received \today} 
\thanks{ This work was supported by the National Key R\&D Program of China nos. 2020YFA0712200, National Natural Science Foundation of China under grant nos. 12226008, 81971611, Shenzhen Science and Technology Program under grant no. RCYX20210609104444089, JCYJ20220818101205012.(Corresponding author: Yanjie Zhu.)}
\thanks{Wenxin Wang and Zhuo-Xu Cui contributed equally to this study.}
\thanks{ Wenxin Wang is with the Shenzhen Institute of Advanced Technology, Chinese Academy of Sciences, Shenzhen 518055, China, and also with the University of Chinese Academy of Sciences, Beijing 101400, China (e-mail: wx.wang1@siat.ac.cn). }
\thanks{ Zhuo-Xu Cui and Dong Liang are with Research Center for Medical AI, Shenzhen Institute of Advanced Technology, Chinese Academy of Sciences, Shenzhen 518055, China(e-mail:{zx.cui; dong.liang}@siat.ac.cn).}
\thanks{ Chentao Cao, Xi Xu, Haifeng Wang and Yanjie Zhu are with Paul C. Lauterbur Research Center for Biomedical Imaging, Shenzhen Institute of Advanced Technology, Chinese Academy of Sciences, Shenzhen 518055, China (e-mail:{ct.cao; xi.xu; hf.wang1; yj.zhu@siat.ac.cn}).}
\thanks{Yanjie Zhu is also with the National Center for Applied Mathematics Shenzhen (NCAMS), Shenzhen 518000, China.}
\thanks{ Yulong Qi, Ziwei Liu and Guanxun Cheng are with Peking University Shenzhen Hospital, Shenzhen, 518036, China (e-mail: 570020720@qq.com; 1448226000@qq.com;18903015678@189.cn). }}

\maketitle

\begin{abstract}
Accurate detection and segmentation of brain tumors is critical for medical diagnosis. However, current supervised learning methods require extensively annotated images and the state-of-the-art generative models used in unsupervised methods often have limitations in covering the whole data distribution. In this paper, we propose a novel framework \textbf{T}wo-\textbf{S}tage \textbf{G}enerative \textbf{M}odel (TSGM) that combines Cycle Generative Adversarial Network (CycleGAN) and \textbf{V}ariance \textbf{E}xploding stochastic differential equation using \textbf{j}oint \textbf{p}robability (VE-JP) to improve brain tumor detection and segmentation. The CycleGAN is trained on unpaired data to generate abnormal images from healthy images as data prior. Then VE-JP is implemented to reconstruct healthy images using synthetic paired abnormal images as a guide, which alters only pathological regions but not regions of healthy. Notably, our method directly learned the joint probability distribution for conditional generation. The residual between input and reconstructed images suggests the abnormalities and a thresholding method is subsequently applied to obtain segmentation results. Furthermore, the multimodal results are weighted with different weights to improve the segmentation accuracy further. We validated our method on three datasets, and compared with other unsupervised methods for anomaly detection and segmentation. The DSC score of 0.8590 in BraTs2020 dataset, 0.6226 in ITCS dataset and 0.7403 in In-house dataset show that our method achieves better segmentation performance and has better generalization.

\end{abstract}


\raggedbottom 

\section{Introduction}
\IEEEPARstart{T}{he} detection of brain tumors, such as glioblastoma\cite{hanif2017glioblastoma}, is crucial for clinical diagnosis and treatment. Magnetic resonance imaging (MRI) plays a pivotal role in brain tumors detection. It has multiple modalities, such as T1-weighted (T1w), T2-weighted (T2w), T1-Weighted Contrast-Enhanced (T1ce), and Fluid Attenuated Inversion Recovery (Flair) MRI images, to provide complementary information of the tumors \cite{tschuchnig2022anomaly,pang2021deep}. However, segmentation of brain tumors is still challenging since they are often characterized by fuzzy boundaries and variabilities in morphology and location. Manual annotation is time-consuming and labor-intensive, leading to interobserver variability and subjective interpretations. Therefore, there is an urgent need for accurate automatic detection and segmentation of brain tumors. 

In recent years, deep learning-based methods \cite{zhou2019intracranial, federau2020improved, havaei2017brain} have been widely used for anomaly detection and segmentation of medical images. These methods can be generally divided into two categories: supervised and unsupervised methods. Typical supervised methods employ CNN-based networks to learn the mapping from the input image to the corresponding label (tumor region) through training data or prior information \cite{liu2020multi,alijamaat2021multiple,lachinov2019glioma,kaur2021ga}. These methods leverage the capabilities of CNNs in feature extraction and nonlinear fitting, making them well-suited for brain tumor analysis. However, supervised methods need diverse annotated training samples, which are costly to obtain. Their extensive usage is limited since they can only detect already-learned specific pathologies \cite{han2021madgan}.

Given the above limitations, unsupervised anomaly detection (UAD) has emerged as a promising alternative. The main principle for almost all UAD studies is using the discrepancy between the anomalous image and its corresponding healthy reconstruction obtained from a generative model, referred to as reconstruction-based methods \cite{gong2019memorizing}. Several deep generative models, including Variational Autoencoders (VAEs) \cite{kingma2013auto, uzunova2019unsupervised, chen2018unsupervised}, Generative Adversarial Networks (GANs) \cite{NIPS2014_5ca3e9b1, schlegl2019f, lambert2021leveraging} and diffusion models \cite{ho2020denoising, song2019generative, song2020score,wolleb2022diffusion,pinaya2022fast,sanchez2022healthy,wyatt2022anoddpm} have been applied in UAD to estimate healthy distributions. VAEs utilize the encoder to sample from the encoded latent space and employ the decoder to generate new data. But it often suffers from the blurry issue because the assumption of a normal distribution for the latent variable is not sufficiently complex to accurately capture the true posterior distribution \cite{cai2019multi}. GANs leverage generative and discriminative models to achieve realistic sample generation. However, GAN-based models are challenging to train and may suffer from mode collapse or instability. Both of them can’t cover the whole distribution. 

Rethinking image bidirectional translation using Cycle Generative Adversarial Network (CycleGAN) \cite{zhu2017unpaired}, it can be found that mapping from healthy to diseased images is generally well-posed. In contrast, generating a healthy image from a given diseased input admits multiple unstable solutions, rendering it ill-posed. Therefore, while CycleGAN is often capable of learning the mapping from healthy to diseased images effectively, it struggles to adequately represent the mapping from diseased to healthy images. To address the ill-posed nature of diseased to healthy image translation, regularization becomes a necessary approach. Recently, diffusion models have shown remarkable ability to generate high-quality images and effectively learn data distributions \cite{dhariwal2021diffusion}. Their approach of regularized maximum likelihood estimation is one of the effective methods for addressing ill-posed problems. The diseased image restoration problem can be formulated as the task of estimating the distribution conditioned on the input image. Fundamentally, they transform the original data distribution into a Gaussian distribution in the forward process and then recover the data from the noise by reversing the process, resulting in a generated image. Diffusion models have already been applied in brain anomaly detection. For example, Julian Wyatt et al. \cite{wyatt2022anoddpm} trained a diffusion model named AnoDDPM on healthy samples to map anomalous area onto the healthy distribution. They also show that using Simplex noise over Gaussian noise can improve the performance. Pinaya, Walter HL et al. \cite{pinaya2022fast} utilized the diffusion model to repair the latent abnormal area and decoded the latent space back to the pixel space using VQ-VAE to obtain the pixel-level residual. However, the original diffusion models produce random samples, resulting in uncertainty of the reconstruction results. Therefore, conditional diffusion guidance methods have been proposed to address this issue by adding a gradient or implicit gradient guidance in the diffusion process, such as the classifier and classifier-free guidance. In classifier-guided generation, Julia Wolleb et al. \cite{wolleb2022diffusion} used DDIM to realize image-to-image translation between diseased and healthy subjects by combining the deterministic iterative noising and denoising scheme with the gradient of an additional classifier. Building upon this work, Pedro Sanchez et al. \cite{pinaya2022fast} further improved its performance with the classifier-free guidance. They trained on conditional and unconditional objectives by randomly dropping a certain probability of conditioning information during training to manipulate the generative process instead of classifiers. The above conditional generation exhibits limitations in terms of complexity, dependence on training data, and generalization, among other aspects.

\begin{figure*}
    \centering
    \includegraphics[width=0.9\textwidth]{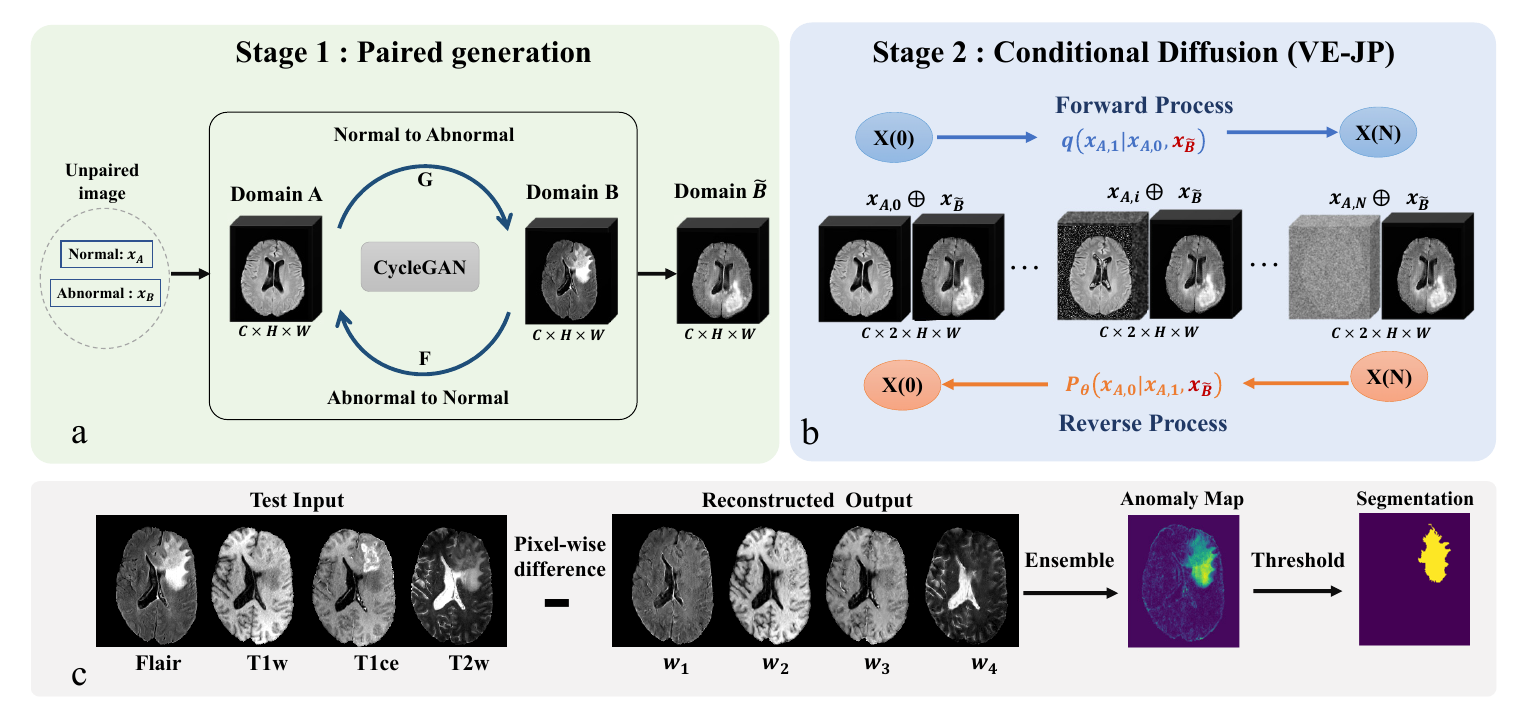}
    \caption{\textbf{Overview of the proposed method.} CycleGAN was used to achieve the transition from domain A(normal) to domain B(abnormal) (stage 1). Fuse the original A and the synthetic $\tilde{B}$:$\{A \cup \tilde{B}\}$ as the input of VE-JP (stage 2). The diffusion model only learned the distribution of A and used the synthetic $\tilde{B}$ as a guidance to remain the regions healthy.}
    \label{fig:fig1}
\end{figure*}

In this paper, we propose a novel framework comprising \textbf{T}wo-\textbf{S}tage \textbf{G}enerative \textbf{M}odel (TSGM) for UAD of brain tumors. Specifically, the first stage leverages CycleGAN to generate paired abnormal (unhealthy) and normal (healthy) images, since obtaining paired healthy and diseased MR images is impractical in reality. In the second stage, the \textbf{V}ariance \textbf{E}xploding stochastic differential equation using \textbf{j}oint \textbf{p}robability (VE-JP) is implemented to learn normal data distributions conditioned on pseudo-paired abnormal images. The residuals between the generated health samples and the original images indicates the abnormalities. Since multi-modality MR images present supplementary information for brain tumor, we assign different weights to each modality to improve the detection accuracy. TSGM was trained on the BraTs2020 brain tumor dataset and was evaluated on two public datasets, as well as an in-house dataset. We also compared its performance to other reconstruction-based anomaly detection benchmarks.  The source code and implementation details are available at
\href{https://github.com/zhyjSIAT/A-Two-Stage-CycleGAN-VE-BRATS2020}{https://github.com/zhyjSIAT/ A-Two-Stage-CycleGAN-VE-BRATS2020}.

The main contributions of our work are highlighted as follows:

\begin{enumerate}
\item We proposed a novel Two-Stage framework for unsupervised anomaly detection. It combines CycleGAN and VE-JP to approximate the conversion process from abnormal to healthy domain and utilizes the difference to detect anomalies effectively. 
\item We devised a new conditional generation of diffusion model called VE-JP. It doesn’t require additional labelled images or gradient trainers but achieves learning the joint distribution of pseudo-paired images to act as a guidance.
\item We introduced an efficient multi-modality MRI ensemble strategy, which combines the results from the various modalities with different weighting to enrich the information associated with each pixel of the brain MRI. 
\end{enumerate}


\section{Methods}
Figure~\ref{fig:fig1} shows the framework of the Two-Stage Generative Model that utilizes CycleGAN and VE-JP. The details of each stage are described in the following.

\subsection{Stage 1 – Paired Image Generation}
Stage 1 generates paired normal and abnormal samples using a CycleGAN, which is trained on unpaired images from either different patients or the same patient. It consists of two separated mapping functions to construct the cycle, i.e., Generator1: normal (A) → abnormal (B) and Generator2: abnormal (B) → normal (A). Both mapping functions are parameterized with identical neural networks. By simultaneously optimizing cycle consistency loss and adversarial loss, CycleGAN achieves unsupervised learning of mapping relationships for image translation. Additionally, it enforces a bijection between the two domains.

The adversarial loss \cite{zhu2017unpaired} is employed to encourage the generator to generate visually similar images to images from the target domain. It is formulated as follows: 

\begin{equation}
\begin{aligned}
    \mathcal{L}_{\mathrm{GAN}}\left(G, D_B, A, B\right)
    &=\mathbb{E}_{b \sim B}\left[\log D_B(b)\right]+\mathbb{E}_{a \sim A}
    \\ \left[\log \left  (1-D_B(G(a))\right]\right].
\end{aligned}
\end{equation}

where G tries to generate images G(a) that closely resemble images from domain B with the discriminator $D_B$.

The assumption of cycle-consistency guarantees that the translated abnormal images retain similar texture information as the target domain B while disregarding any potential geometric alterations. The cycle consistency loss is defined as follows:

\begin{equation}
\begin{aligned}
    \mathcal{L}_{\mathrm{cyc}}(G, F)=\mathbb{E}_{a \sim A}\left[\|F(G(a))-a\|_1\right]+\mathbb{E}_{b \sim B}
    \\ \left[\|G(F(b))-b\|_1\right].
\end{aligned}
\end{equation}

Therefore, the loss of CycleGAN is 

\begin{equation}
\begin{aligned}
    \mathcal{L}\left(G, F, D_A, D_B\right)=\mathcal{L}_{\mathrm{GAN}}\left(G, D_B, A, B\right)+
    \\ \mathcal{L}_{\mathrm{GAN}}\left(F, D_A, B, A\right)+\lambda \mathcal{L}_{\mathrm{cyc}}(G, F).
\end{aligned}
\end{equation}

Here coefficient $\lambda$ governs the relationship between the adversarial and cycle consistency losses. 

As mentioned above, the source of healthy input is $x_A$, and the corresponding abnormal image synthesized by CycleGAN is denoted as $x_{\tilde{B}}$ . $x_A$ and $x_{\tilde{B}}$ are of dimension (C, H, W), where C denotes the number of channels, and (H, W) denotes the height and width of the images.

\subsection{Stage 2 - Condition Diffusion}

Stage 2 is to generate healthy images using a conditional diffusion model. With the pseudo-paired data generated by Stage 1, VE-JP model learn the joint distribution $p\left( x_A | x_{\tilde{B}} \right)$ using the score-matching method for healthy image generation. Figure~\ref{fig:fig1}(b) illustrates the framework of the VE-JP model. 

We follow the implementation of Variance Exploding (VE) SDE in the score-based generative models (SGMs) \cite{song2020score}.
In the classical VE-SDE, the key idea is to model the gradient of the log probability density function of $x_A$, a quantity is often known as the score function $s_{\theta^{*}}(x,i) \approx \nabla_{x_A} \log p_i(x_A)$ of $x_A$, where $x_A$ is the only input. For the conditional generation, the score function becomes \cite{taofeng2022brain}:

\begin{equation}
\begin{aligned}
    s_{\theta^*}\left(x_A, x_{\tilde{B}}, \mathrm{i}\right) & \approx \nabla_{x_A} \log p_i\left(x_A \mid x_{\tilde{B}}\right) \\
& =  \nabla_{x_A}\left(\log p_i\left(x_A, x_{\tilde{B}}\right)-\log p_i\left(x_{\tilde{B}}\right)\right) \\
& =\nabla_{x_A} \log p_i\left(x_A, x_{\tilde{B}}\right).
\end{aligned}
\end{equation}

The final simplification of the gradient $\nabla_{x_A} \log p_i\left(x_A, x_{\tilde{B}}\right)$ can be transformed into directly training the diffusion model with the joint input $\left(x_A, x_{\tilde{B}}\right)$ in the channel dimension to achieve conditional controlled generation.

In the forward process of VE-JP, a small amount of Gaussian noise is added to the healthy image $x_A$ to perturb the data and gradually transform into a Gaussian distribution. Its paired abnormal image $x_{\tilde{B}}$ are fixed (as shown in Alg.\ref{alg1}). The structure of healthy data is progressively degraded with a fixed Markov chain over 1000 steps as:

\begin{equation}
\begin{aligned}
    X_{i+1}\left(x_{A, i+1}, x_{\tilde{B}}\right)=X_i\left(x_{A, i}, x_{\tilde{B}}\right)+ 
    \\ \sqrt{\sigma_{\min }^2\left[\left(\frac{\sigma_{\max }}{\sigma_{\min }}\right)^{2 i}-I\right]} z_i, \; i=1,2, \ldots, N.
\end{aligned}
\end{equation}

Where $X_i$ is the $i^{th}$ joint perturbed data $(X_i(x_{A,i},x_{\tilde{B}} ))$, defined by $X_i:=x_{A,i} \bigoplus x_{\tilde{B}}$. $\bigoplus$ concatenates the input $x_A$ and $x_{\tilde{B}}$ , so $X_i$ is of dimension (C, 2, H, W). $\left\{\sigma_i\right\}_{i=1}^N$ is the sequence of noise scales $\sigma_{\min }=\sigma_1<\cdots<\sigma_N=\sigma_{\max }$. Specifically, $\sigma_{min}$ is small enough such that  $p_{\sigma_{\text {min}}}(X) \approx p_{\text {health}}\left(X_0\left(x_{A, 0}, x_{\tilde{B}}\right)\right)$, and $X_0$ is the joint distribution of the normal image and synthetic abnormal image. ${\sigma_{max}}$ is large enough such that $p_{\sigma_{\max }}(X) \approx N\left(X_N\left(x_{A, N}, x_{\tilde{B}}\right);0,\sigma_{\max}^2I\right)$, and $X_N$ denotes the joint distribution of standard Gaussian distribution for $x_{A,N}$ and $x_{\tilde{B}}$ at the moment N. 

During the reverse diffusion process, the healthy image can be recovered from the noise based on a reverse-time SDE derived from the forward SDE with the guidance of abnormal image $x_{\tilde{B}}$. The reverse-time SDE can be rewritten as:

\begin{equation}
\begin{aligned}
   \mathrm{dx}=\left[f(\mathbf{x}, i)-g(i)^2 s_{\theta^*}\left(x_A, x_{\tilde{B}}, i\right)\right] \mathrm{d} i+g(i) \mathrm{dw}.
\end{aligned}
\label{eq:reverse}
\end{equation}

where w represents a standard Wiener process as time progresses in reverse from N to 0, and $\mathrm{d}i$ represents an infinitesimal negative timestep. Once the scores of each marginal distribution for all t are known, we can derive the reverse diffusion process from Eq. (\ref{eq:reverse}) and simulate it to obtain healthy samples from abnormal images.

The time-dependent score function $s_{\theta^*}(X_{i+1},i+1), i\in {1,2,\ldots,N}$ is trained on a U-Net with the objection function as :

\begin{equation}
\begin{aligned}
    L(\theta ; \sigma)=\frac{1}{2} \mathbb{E}_{p_i(X)} [||\sigma_{\min }^2\left[\left(\frac{\sigma_{\max }}{\sigma_{\min }}\right)^{2 i}-\mathrm{I}\right]
    \\  s_{\theta^*}\left(X_{i+1}, \sigma_{\min }\left(\frac{\sigma_{\max }}{\sigma_{\min }}\right)^i\right)+z||_{2}^2 ].
\end{aligned}
\end{equation}

The Predictor-Corrector method framework (PC Sampling) is used to generate samples \cite{song2020score} (as shown in Alg.\ref{alg2}). Predictor and corrector are executed alternately. The Predictor is reverse diffusion (from the joint distribution of noise $x_N$ and $x_{\tilde{B}}$ to the distribution of $x_A$) and the corrector is Langevin dynamics. 

\begin{algorithm}
    \DontPrintSemicolon
    \caption{Training}\label{alg1}
    \SetAlgoLined
    \LinesNumbered
    \textbf{repeat} \;
    $\left( x_{A,0},x_{\tilde{B}}\right) \sim p \left( x, x_{\tilde{B}}\right)$ \;
    $ i \sim \mathcal{U}(\mathbf{0}, \mathbf{1})$ \;
    $\mathbf{z} \sim \mathcal{N}(\mathbf{0}, \mathbf{I})$ \;
    $X_{i+1}\left(x_{A, i+1}, x_{\tilde{B}}\right)=X_i\left(x_{A, i},x_{\tilde{B}}\right)+\sqrt{\sigma_{\min }^2\left[\left(\frac{\sigma_{\max }}{\sigma_{\min }}\right)^{2 i}-I\right]} \mathbf{z}_i$ \;
    Take a gradient descent step on
    $\nabla_\theta\left\|s_\theta\left(x_{A, i}, x_{\tilde{B}}, i\right)+\mathbf{z}\right\|^2$ \;
    \textbf{until} converged
\end{algorithm}

\begin{algorithm}
    \caption{Anomaly detection and segmentation using VE-JP}\label{alg2}
    \DontPrintSemicolon
    \SetAlgoLined
    \LinesNumbered
    \textbf{Input:}{ the healthy image $x_A$, the synthetic paired abnormal image $x_{\tilde{B}}$ , N, K, m}\;
    \textbf{Output:} {the $x_{A,0}$, the anomaly map and segmentation mask}
    $x_{A,N} \sim \mathcal{N}(\mathbf{0}, \sigma_{\text{max}}^2 \mathbf{I}) $ \;
    $x_{N} \leftarrow x_{A,N} \bigoplus x_{\tilde{B}} $ \;
    \For{$i=N-1$ to $1$}{
      $ X_i^{\prime} \leftarrow X_{i+1}+\left(\sigma_{i+1}^2-\sigma_i^2\right) s_{{{\boldsymbol{\theta}}}^*}\left(X_{i+1}, \sigma_{i+1}\right)$ \;
      $\mathbf{z} \sim \mathcal{N}(\mathbf{0}, \mathbf{I})$ \;
      $X_{i+1}=X_i^{\prime}+\sqrt{\sigma_{i+1}^2-\sigma_i^2} \mathbf{z}$ \;
      \For{$j \leftarrow 1$ to $K$}
        {
          $\mathbf{z} \sim \mathcal{N}(\mathbf{0}, \mathbf{I})$ \;
          $\mathbf{g}  \leftarrow s_{{{\boldsymbol{\theta}}}^*}(x_i^{j-1}, \sigma_i)$ \;
          $\epsilon  \leftarrow 2\left(r\|\mathbf{z}\|_2 /\|\mathbf{g}\|_2\right)^2 $ \;
          $X_i^j  \leftarrow X_i^{j-1}+\epsilon_i \mathbf{g}+\sqrt{2 \varepsilon_i} \mathbf{z}_i$ \;
        }
        $X_{i-1}^0 \leftarrow X_{i}^K$\;
    }
    $\text{weighted\_{anomap}} = \sum_{v=1}^{m} w_v\left(x_{Av} - x_{Av}^0\right) $ \;
    $\text{mask} = \text{thresholding(weighted\_{anomap})}$ \;
 
    \Return $X_A^0$, $weighted_{anomap}$, segmentation mask;
    
\end{algorithm}

\subsection{Multi-modality MRI Ensemble}

The ensemble process uses the pixel-level fusion of the different modalities governed by Eq.(\ref{eq:ensemble}). 

\begin{equation}
\begin{aligned}
   I_{\text{diff\_ensemble}} = \sum_{v=1}^{m} \left( w_v(x_{Av} - x_{Av}^0) \right), s.t. \sum_{v=1}^{m} w_v = 1.
\end{aligned}
\label{eq:ensemble}
\end{equation}

where $x_{Av}$ is a single modality image obtained by VE-JP, and $x_{Av}^0$ denotes the corresponding original input. The weight $w_v$ corresponds to the credibility of each modality, and the sum of total weight equals 1. The difference between the original and reconstructed images is weighted along the channel dimension to obtain the anomaly heatmap, which is subsequently utilized to obtain the segmentation mask through thresholding. For different modalities of the same person, we enumerated multiple sets of weights varying from 0.0 to 1.0 to find the optimal weight for the best performance. The final weight is obtained by computing the average of the optimal weights across all training sets.

\section{Experiments}
In this section we present the datasets used in the experiments, along with the preprocessing methods applied and provide the implementation details and the evaluation metrics.
\subsection{Dataset}


\begin{table*}[!ht]
\scriptsize
\centering
\caption{{\bf Dataset information}. GBM stands for Glioblastoma multiforme, while MBT stands for multiple brain tumors. }
\label{tab1}
\begin{tabular}{c c p{2.5cm}<{\centering} c c c}
  \toprule
  \multirow{2}{*}{Parameter} & \multicolumn{2}{c}{Public Datasets} & \multicolumn{3}{c}{Inhouse Dataset} \\
  \cmidrule(lr){2-3} \cmidrule(lr){4-6}
  & BraTs  & ICTS  & \multicolumn{3}{c}{In-house} \\
  & (GBM) \cite{menze2014multimodal,bakas2017advancing,bakas2018identifying} & (MBT) \cite{lu2021intracranial,dorjsembe2022three} & \multicolumn{3}{c}{(GBM)} \\
  \hline
  Training & 332 & 0 & \multicolumn{3}{c}{0} \\
  Testing & 37 & 192 & \multicolumn{3}{c}{50}  \\
  \textbf{Image parameters}\\
    Scanner	& Diverse &	GE Signa Excite 1.5 T and Siemens Aera 1.5 T &	\multicolumn{3}{c}{UIH uMR780 3.0T} \\
    Modality & T1w T1ce T2w Flair & T1ce & T1-Flair T1-Flair-ce	& T2-Flair &  T2w  \\
    Resolution(mm3) & 1×1×1 & 1×1×1 & 0.598×0.598×4.0 & 0.599×0.599×4.0 & 0.456×0.456×4.0 \\
    In-plane size & 240×240 & 192×192 & 334×384 & 300×384 & 438×504 \\
    Slices & 155 &192 & \multicolumn{3}{c}{$25 \sim 26$} \\
    Skull-stripped & No & HD-BET & \multicolumn{3}{c}{HD-BET} \\
    Co-registered & No & SRI24 & \multicolumn{3}{c}{SRI24}\\ 
  \bottomrule
\end{tabular}
\label{tab:datasets}
\end{table*}

\textbf{BraTs2020} \cite{menze2014multimodal,bakas2017advancing,bakas2018identifying} The dataset is consisted of glioma images provided by Multimodal Brain Tumor Segmentation Challenge (BraTs2020). The data were collected from 19 institutions using various MRI scanners. 369 patients were involved and four modalities (T1w, T1ce, T2w, Flair) were acquired for each patient. Those images have rigidly aligned, interpolated to the isotropic resolution of $1\times 1 \times 1 mm^3$, and skull-stripped. The in-plane matrix size is fixed at $240 \times 240$, and the slice number is 155 with pixel-level annotations for the lesions in each patient. The initial ground truth labels include four classes : background, GD-enhancing tumor, the peritumoral edema, and the necrotic and non-enhancing tumor core.

\textbf{ICTS}\cite{lu2021intracranial,dorjsembe2022three} The dataset is provided by the Intracranial Tumor Segmentation (ICTS) Challenge, encompassing a diverse range of brain tumors that includes meningiomas, schwannomas, pituitary adenomas, etc. The data were collected by National Taiwan University Hospital on GE Signa Excite 1.5 T and Siemens Aera 1.5 T scanners. This train-2207 dataset comprises 192 labeled T1ce MR images and was only used for testing. The images have an in-plane matrix size of $192 \times 192$, with 192 slices. The provided labels are similar to the tumor core of BraTS2020.

\textbf{In-house} The research use of the In-house dataset has been approved by the Ethics Committee of Peking University Shenzhen Hospital (Ethics Approval (Research) [2022] No.107). The preoperative images of 50 patients with glioma were collected. They were acquired on a 3T MR scanner (United Imaging Healthcare) between 2022 and 2023. The images were de-identified before analysis to protect patient privacy. Images in T1-Flair T1-Flair-ce T2-Flair and T2w modalities have different resolutions and sizes. The segmentation labels in axial-slice level were provided by a radiologist with more than 15 years of experience. Detailed information of all datasets is listed in Table \ref{tab:datasets}.

\subsection{Data Preprocessing}

For BraTs dataset, we shuffled the training set and split it into two subsets with a ratio of 9:1. Slices without identified tumor on the ground truth mask are defined as healthy slices. We sliced the 3D MR scans into axial slices and chose 80 to 128 slices out of the original 155 slices, as tumors are rarely on the upper and lower regions of the brain. Additionally, we discarded any blank or redundant slices with pixel values less than 15. We have merged the three tumor classes into one combined class. Therefore, we refined the segmentation task to proficiently discern between tumor and background regions. More specifically, our training set consisted of 5,795 healthy slices and 10,473 diseased slices, while the testing set included 508 healthy and 1282 diseased slices. Since each slice contained four modality images, we used a total of 23,180 healthy images and 41,892 diseased images for training. The volume of slices was padded to $256\times 256$ to satisfy the input layer requirements of the model. 

For ICTS and In-house dataset, the applied pre-processing routines include conversion of the DICOM files to the NIFTI file format, brain extraction with HD-BET \footnote{https://github.com/MIC-DKFZ/HD-BET}, co-registration to the same anatomical template (SRI24) using $\text{brain\_sas\_baseline}$ \footnote{https://github.com/FeliMe/\text{brain\_sas\_baseline}}, crop their slice to $70\sim120$. Furthermore, pixel intensity normalization is the same as the BraTs process by cutting the top and bottom one percentile of pixel intensities before applying min-max normalization.

\subsection{Implementation details}

\textbf{Benchmarks.} The proposed method compared with five benchmark methods: (i) DDIM with classifier guidance ($\text{DPM\_classifier}$) \cite{wolleb2022diffusion}; (ii) DDIM with classifier-free guidance ($\text{DPM\_classifier\_free}$) \cite{sanchez2022healthy}; (iii) CycleGAN \cite{zhu2017unpaired}; (iv)the variational autoencoder (VAE) 
\cite{schlegl2017unsupervised}, (v) Fast unsupervised anomaly detection with generative adversarial networks (f-anogan) \cite{schlegl2019f}, with the results of each method tuned to the best \cite{schlegl2017unsupervised}. 

\textbf{Setting.} One of the major distinctions in the training setup is the utilization of channel dimensions. Except for our method and CycleGAN, other methods concatenate all four modalities at the channel dimension for each patient during training. Our experiments were conducted using PyTorch 1.7.0 as the software framework and 4 Nvidia A100 GPUs for model training. For CycleGAN, we employed the $\text{resnet\_9blocks}$ as the generator and the PatchGAN \cite{isola2017image} as the discriminator. For VE-JP, the BigGAN residual block was used to perform both upsampling and downsampling of the activations during training \cite{dhariwal2021diffusion}. Our model was trained with 50 epochs and the noise schedule N=1000 steps. The exponential moving average (EMA) rate was 0.999. To control the noise level in forward diffusion, VE-JP was configured with $\sigma_{max}=348$ and $\sigma_{min}=0.1$ \cite{song2020improved}. In addition, the other models were trained with a learning rate of $10^{-4}$ using the Adam optimizer. Both $\text{DPM\_classifier}$ and $\text{DPM\_classifier\_free}$ models maintained consistent hyperparameter settings. Specifically, the DDPM model utilized a hybrid loss function and a sampling step of T = 1000, with the remaining hyperparameter detailed in the appendix of \cite{dhariwal2021diffusion}. For f-anogan and VAE, the z-dimension was set at 128 with parameter configurations specified in \cite{baur2021autoencoders}. 

\subsection{Evaluation Metrics}

\begin{figure*}
    \centering
    \includegraphics[width=0.7\textwidth]{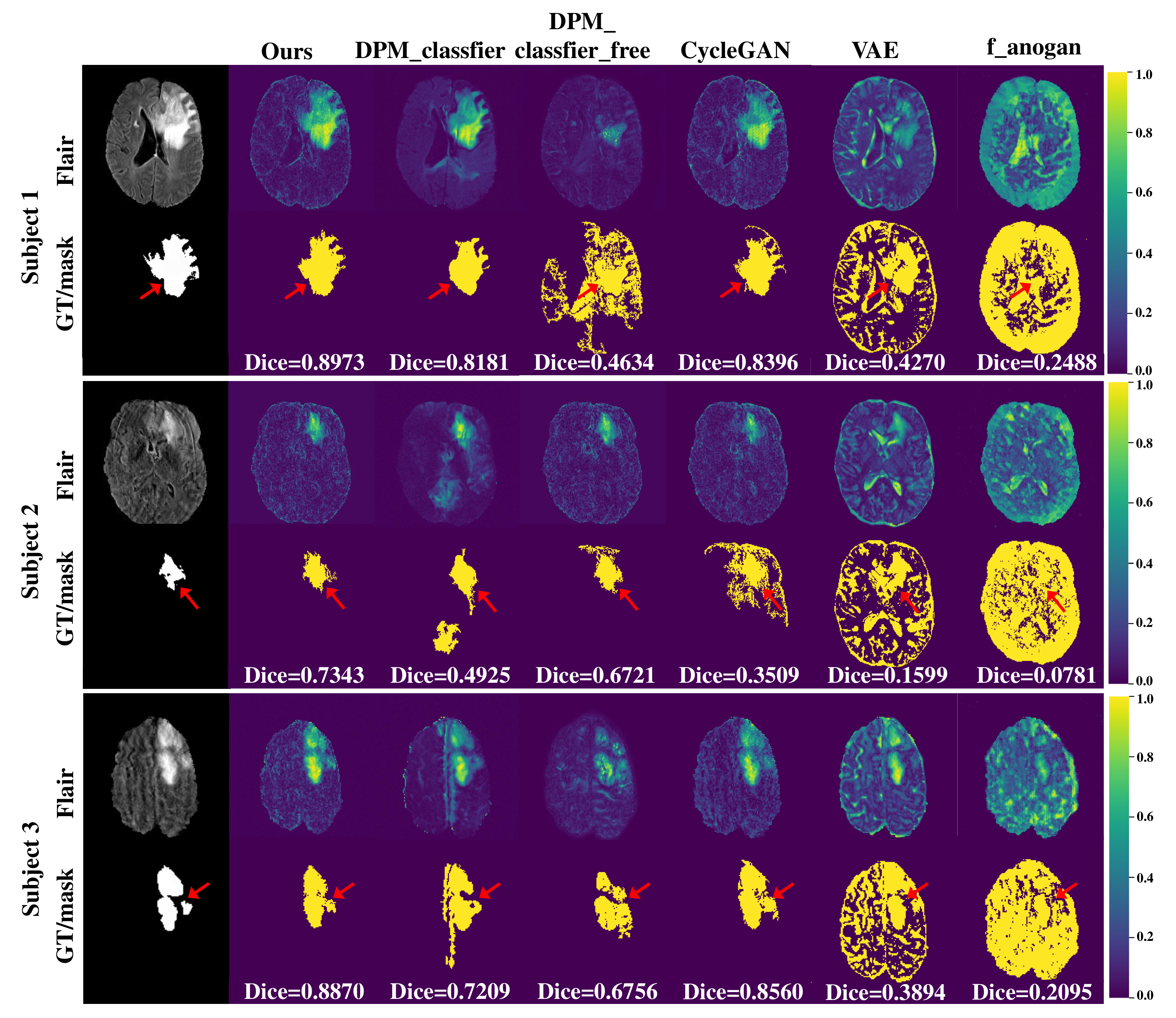}
    \caption{\textbf{Representative results from BraTs 2020 (Patients no.6, 34 and 39, respectively).} The first column indicates the input image (Four MRI modalities were used in training, but here we only show the Flair image) and the ground truth masks. Other columns show the anomaly maps and masks generated by the different benchmarks.}
    \label{fig:fig2}
\end{figure*}


For quantitative assessment, six metrics were used to evaluate the performance of detection and segmentation with our method. 

\begin{itemize}
  \item Dice similarity coefficient (DSC): The degree of overlap between the predicted mask and ground truth labels. 

  \begin{equation}
  \begin{aligned}
        DSC = \frac{2 \times TP }{2 \times TP + FN + FP}
  \end{aligned}
  \label{eq:DSC}
  \end{equation}
 where TP and FN is the number of true positives and false negatives, respectively. TN and FP denotes the number of true negatives and false positives, respectively.
  \item Precision (PRE) and Recall (REC): REC is the proportion of actual positive samples that are correctly identified. PRE refers to the proportion of positive identifications was actually correct. 
  \begin{equation}
    \begin{aligned}
        PRE = \frac{TP}{TP + FP} \quad
        REC = \frac{TP}{TP + FN}
    \end{aligned}
  \end{equation}
  \item Area under the precision-recall curve (AUPRC): The area under the curve formed by the PRE corresponding to different True Positive Rates (TPR), where TPR represents the proportion of correctly predicted true positive samples among all actual positive samples. 
  \begin{equation}
    \begin{aligned}
    AUPRC & =\int_0^1 P(\text {recall}) d(\text {recall}) 
    \end{aligned}
  \end{equation}
  Here, $P(\text {recall})$ indicates the Precision value corresponding to each recall value, while $d(\text {recall})$ refers to the slight change in recall value.
  \item Intersection over Union (IoU): The overlap between the predicted and ground truth regions of interest by calculating the ratio of the intersection to the union of the two regions.
 \begin{equation}
    \begin{aligned}
        IoU = \frac{TP}{TP + FN + FP}
    \end{aligned}
 \end{equation}
 \item Hausdorff95 Distance (HD95): The 95th percentile of the Hausdorff distance (HD) which measures the maximum distance between set A and the nearest point in set B.

\begin{equation}
    \begin{aligned}
    	\textbf{H}(A,B) = max\{&h_{(A,B)}, h_{(B,A)}\} \\
    		   = max\{ &max_{a \in A} min_{b \in B} d(a,b), \\
    		   &max_{b \in B} min_{a \in A} d(a,b)\}.
    \end{aligned}
    \label{eq:HD}
\end{equation}

\end{itemize}

Except for HD95, the other metrics range from 0.0 to 1.0, where higher values indicate better performance. All these metrics provide a comprehensive evaluation of both detection and segmentation performance.

\begin{figure*}
    \centering
    \includegraphics[width=0.7\linewidth]{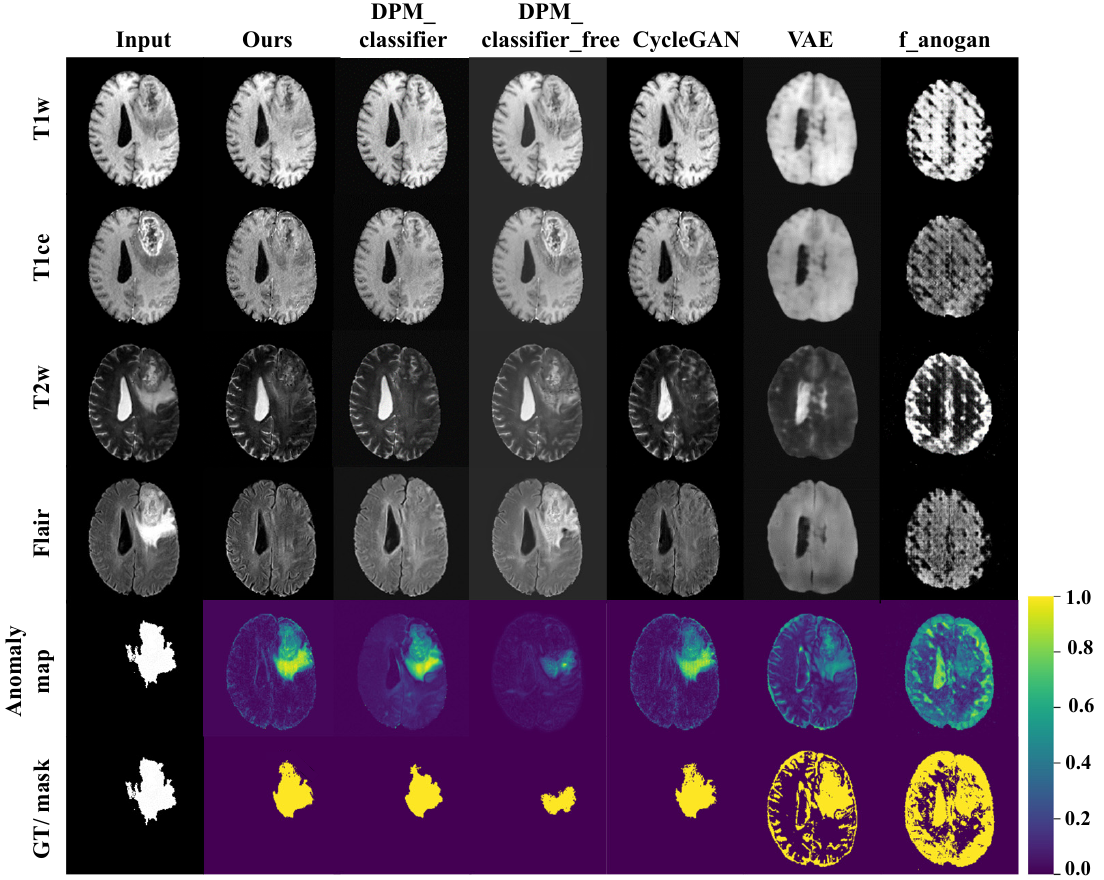}
    \caption{\textbf{Example with ID 006 from the BraTs2020 dataset.} Each subplot presents the reconstruction results, anomaly map and segmentation mask for an image with different methods in each modality which from left to bottom: T1w, T1ce, T2w and Flair.}
    \label{fig:fig3}
\end{figure*}

\section{Results}

To demonstrate the performance of TSGM, we compared it with five other benchmarks. 

\subsection{Anomalous visualization}

Figure~\ref{fig:fig2} displays the anomaly heatmaps and segmentation masks of three representative cases from BraTs compared to other benchmarks. Our proposed method achieves remarkably detailed anomaly maps in various layers of the brain tumor region and for different tumor sizes. Additionally, it outperforms other benchmarks when dealing with irregularly shaped tumors, lesions with ambiguous boundaries or smaller tumors. Notably, as indicated by the red arrows in Fig.~\ref{fig:fig2}, our approach accurately captures abrupt transitions in tumor boundaries, successfully addressing concerns related to over-detection and under-detection observed in other approaches. In subject 3 (the third row of Fig.~\ref{fig:fig2}), we have found that the provided tumor label exhibits a disconnected region, which raises the need for further discussion regarding the accuracy and validity of these labels.

Figure~\ref{fig:fig3} depicts the results under different modalities. It is observed that VAE-based models tend to blur images, which inherently hampers segmentation performance. Furthermore, f-anogan exhibits inaccurate reconstruction image and is inclined to generate anomalous structures. The $\text{classifier\_free\_guided}$ method fails to fully restore the diseased regions. The generated masks obtained by our method are able to identify the tumors and other high frequency details, while the other methods either change normal parts of the image, or are not able to find an anomaly.

\begin{figure*}
    \centering
    \includegraphics[width=0.7\linewidth]{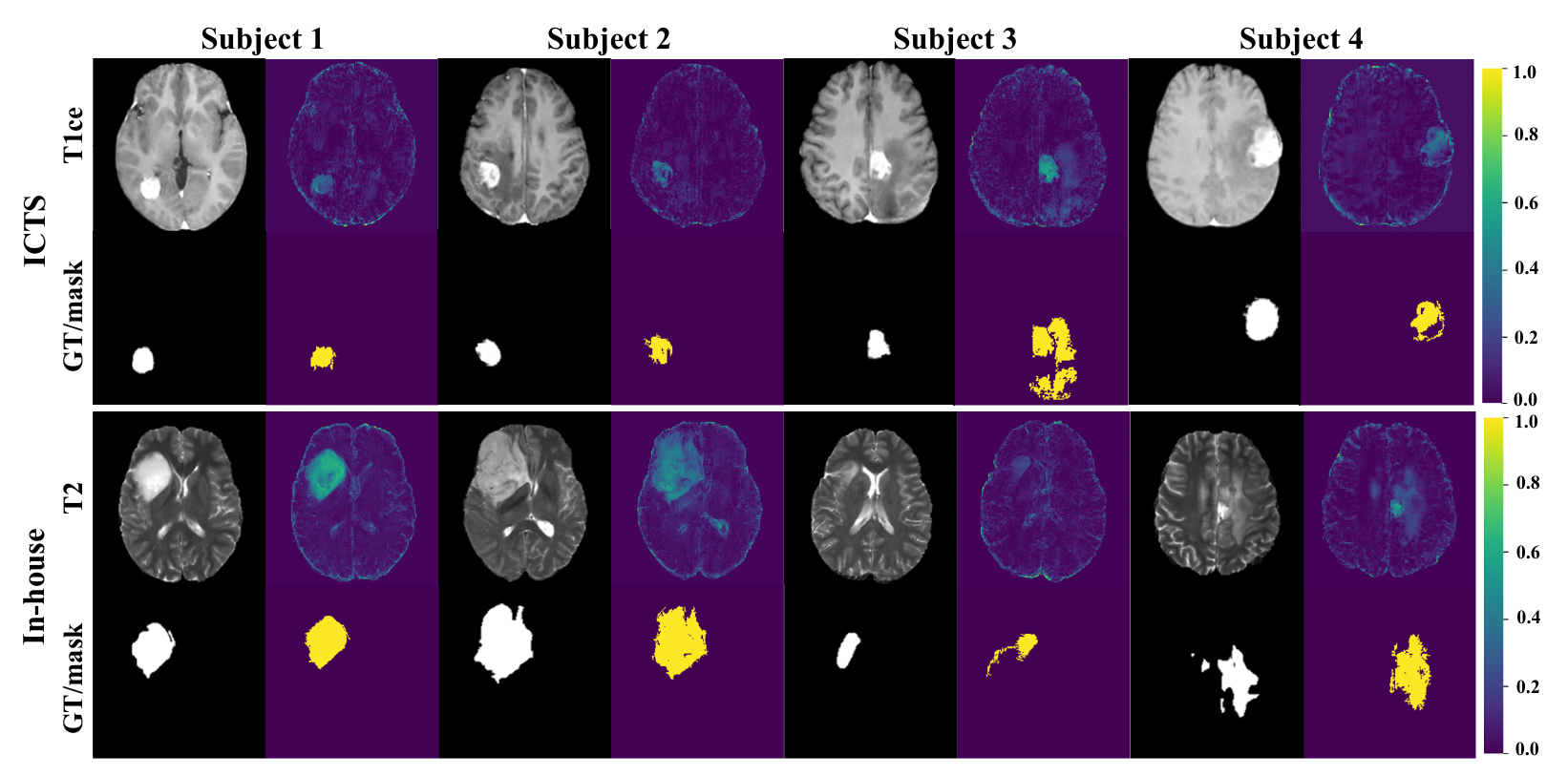}
    \caption{\textbf{Results of the anomalous region localization in the ICTS and In-house dataset.} First and third row: Original image and anomaly map. Second and fourth row: Ground truth and segmentation mask.}
    \label{fig:fig4}
\end{figure*}

Figure~\ref{fig:fig4} depicts the visualization results of anomalous region localization in the ICTS dataset and In-house dataset. As can be observed, our model performs well in detecting tumors of smaller size in different locations, as well as tumors near the brain's edge in the ICTS dataset. Additionally, it demonstrates effective detection of complex edema regions in the In-house dataset. But it is worth noting that the presence of high signal intensity in the T2 modality within the In-house dataset may adversely affect the reconstruction results.

\subsection{Quantitative comparison }

\begin{figure}
    \centering
    \includegraphics[width=1.\linewidth]{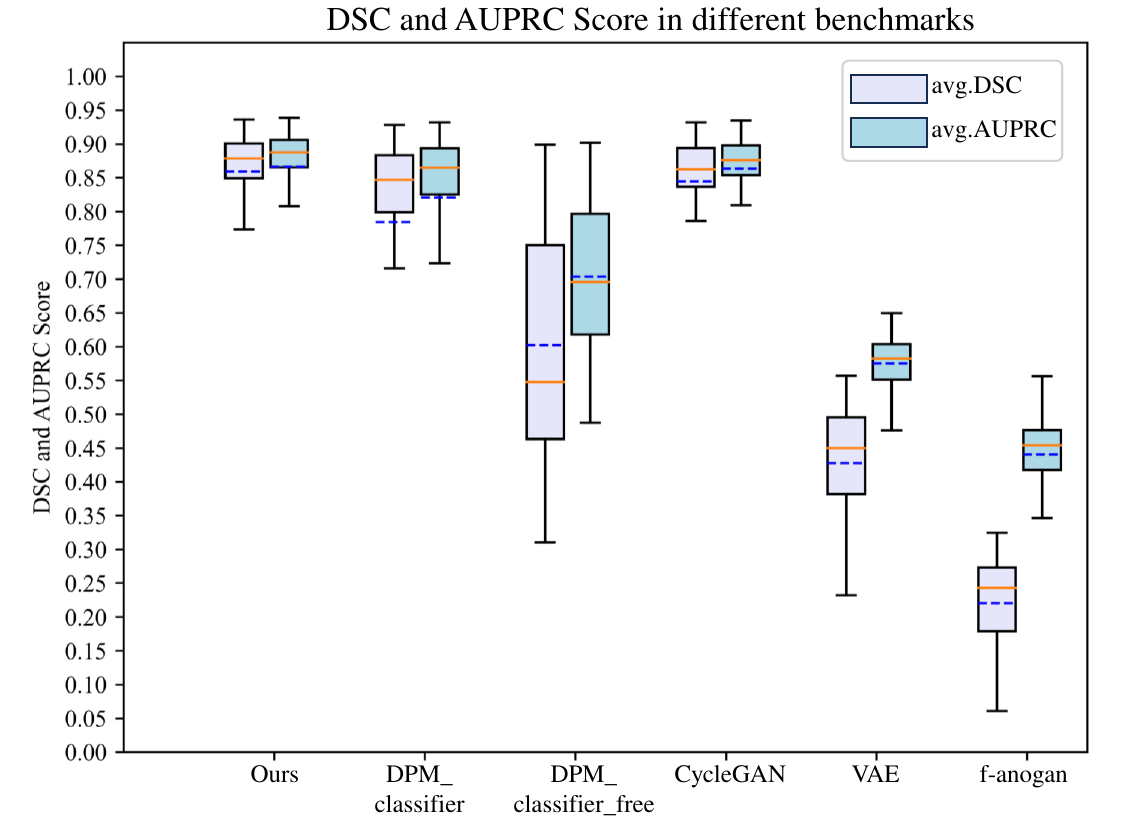}
    \caption{Box plots of the AUPRC and DSC score for the BraTs2020 dataset in different benchmarks. (The solid \textcolor{orange}{orange} line represents the median, while the dashed \textcolor{blue}{blue} line represents the mean.)}
    \label{fig:box}
\end{figure}

For a comprehensive comparison, we employed quantitative metrics to analyze the results of various models in the BraTs dataset. Figure~\ref{fig:box} illustrates the DSC and AUPRC Score box plots tested in different methods. Our proposed TSGM method achieved significantly higher mean values (represented by the \textcolor{blue}{blue} dashed line in Fig.~\ref{fig:box}) of 0.8590 for DSC and 0.8684 for AUPRC compared to the results obtained from other methods.

\begin{table}
\centering
\caption{Anomaly detection performance metrics of different datasets using different methods. \textcolor{red}{Red} indicates the best result, and \textcolor{blue}{blue} displays the second-best. }
\label{tab:overview}
\resizebox{\columnwidth}{!}{
\begin{tabular}{ccccccc}
   \toprule
   \multirow{2}{*}{} & \multicolumn{6}{c}{(Mean \( \pm \) Standard deviation)} \\
   \cmidrule(lr){2-7}
     &  DSC  & AUPRC  & IoU  & PRE  & REC  & HD95\\
    \hline
    BraTs \\
   \hline
   CycleGAN \cite{zhu2017unpaired} & \textcolor{blue}{84.58±0.09} & \textcolor{blue}{86.45 ±0.06} & \textcolor{blue}{74.77±0.15} & \textcolor{blue}{90.16±0.14} & 82.16±0.09 & 7.66±5.56 \\
   VAE \cite{schlegl2017unsupervised} &  42.79±0.08 & 57.54±0.04 & 27.56±0.07 & 29.19±0.07 & \textcolor{red}{85.43±0.07} & \textcolor{blue}{3.38±1.49} \\
   f-anogan \cite{schlegl2019f} & 22.06±0.07 & 44.08±0.06 & 12.57±0.04 & 13.16±0.05 & 74.25±0.13 & \textcolor{red}{2.86±0.77} \\
   {$\text{DPM\_classifier\_free}$ \cite{sanchez2022healthy}} & 60.19±0.17 & 70.32±0.11 & 45.26±0.18 & 77.74±0.29 & 61.57±0.21 & 18.09±16.80 \\
   {$\text{DPM\_classifier}$ \cite{wolleb2022diffusion}} & 78.37±0.19 & 81.95±0.14 & 67.50±0.20 & 84.57±0.26 & 78.64±0.08 & 8.65±5.32 \\
   {TSGM (Ours)} &  \textcolor{red}{85.90±0.12} & \textcolor{red}{86.84±0.09} & \textcolor{red}{75.97±0.11} & \textcolor{red}{91.02±0.09} & \textcolor{blue}{82.44±0.06} & 5.08±3.26 \\
   \hline
   ICTS &  \\
   \hline
   CycleGAN \cite{zhu2017unpaired} & \textcolor{blue}{55.57±0.13} & \textcolor{blue}{55.91±0.15} & \textcolor{blue}{45.35±0.09} & \textcolor{blue}{58.16±0.11} & \textcolor{blue}{53.11±0.24} & \textcolor{red}{9.25±8.11} \\
   VAE \cite{schlegl2017unsupervised} & — & — & — & — & — & — \\	
   
   f-anogan \cite{schlegl2019f}  & — & — & — & — & — & — \\
   {$\text{DPM\_classifier\_free}$ \cite{sanchez2022healthy}} & — & — & — & — & — & — \\
   {$\text{DPM\_classifier}$ \cite{wolleb2022diffusion}} &  31.00±0.29 & 37.24±0.31 & 22.50±0.24 & 30.89±0.35 & 42.07±0.34 & 10.0±0.29 \\
   {TSGM (Ours)} &  \textcolor{red}{62.26±0.17} & \textcolor{red}{66.06±0.18} & \textcolor{red}{53.04±0.13} & \textcolor{red}{62.66±0.15} & \textcolor{red}{65.96±0.13} & \textcolor{blue}{9.86±5.16} \\
    \hline
    In-house \\
   \hline
   CycleGAN \cite{zhu2017unpaired} & \textcolor{blue}{65.34±0.30} & \textcolor{blue}{72.68±0.22} & \textcolor{blue}{55.27±0.31} & \textcolor{blue}{64.32±0.36} &  \textcolor{blue}{80.59±0.16} & \textcolor{blue}{6.75±5.22} \\
   VAE \cite{schlegl2017unsupervised} &  19.92±0.14 & 36.28±0.16 & 11.76±0.09 & 12.63±0.10 & 59.19±0.25 & 7.75±4.28 \\
   f-anogan \cite{schlegl2019f} & 15.03±0.09 & 45.64±0.10 & 8.37±0.05 & 8.51±0.05 & \textcolor{red}{82.46±0.18} & \textcolor{red}{2.39±1.63} \\
   {$\text{DPM\_classifier\_free}$ \cite{sanchez2022healthy}} & 24.28±0.19 & 54.36±0.08 & 15.29±0.14 & 31.22±0.38 & 65.46±0.34 & 8.14±6.93 \\
   {$\text{DPM\_classifier}$ \cite{wolleb2022diffusion}} & 21.87±0.25 & 37.33±0.24 & 14.90±0.19 & 26.68±0.36 & 27.95±0.23 & 15.53±10.16 \\
   {TSGM (Ours)} &  \textcolor{red}{74.03±0.21} & \textcolor{red}{76.72±0.24} & \textcolor{red}{63.73±0.31} & \textcolor{red}{76.62±0.35} &76.31±0.20 & 8.84±4.03 \\
   
   \bottomrule
\end{tabular}
}
\label{tab:methods}
\end{table}


The quantitative metrics on the three datasets using different methods are shown in Table \ref{tab:methods}. It's important to note that the ICTS dataset contains only one modality, T1ce, which means that models such as DPM-classifier-free, f-anogan, and VAE, trained on four modalities, are unable to generate test results for this dataset. The results for DPM-classifier on ICTS were obtained using a single-modality model trained on Flair data. 

Our proposed model maintains strong performance on ICTS and In-house dataset. The suboptimal performance in ICTS may be that the label of ICTS is similar to that of BraTs in the core tumor part, whereas our model detects the whole tumor area including the edema area.

\begin{table}
\centering
\caption{Anomaly detection performance metrics of different input modalities or channels. \textcolor{red}{Red} indicates the best result, and \textcolor{blue}{blue} displays the second-best.}
\label{tab:overview}
\resizebox{\columnwidth}{!}{
\begin{tabular}{ccccccc}
   \toprule
   \multirow{2}{*}{} & \multicolumn{6}{c}{(Mean \( \pm \) Standard deviation)} \\
   \cmidrule(lr){2-7}
     &  DSC  & AUPRC  & IoU  & PRE  & REC  & HD95\\
   \hline
   CycleGAN \cite{zhu2017unpaired} & \\
   \hline
   {Flair(only)} &  70.95±0.24 & 75.66 ±0.16 & 59.18±0.23 & 71.14±0.29 & 78.54±0.12 & 7.14±5.18 \\	
   {T2(only)} & 55.97±0.21 & 60.33±0.21 & 41.58±0.19 & 66.51±0.31 & 52.73±0.18 & 16.21±7.53 \\
   {single-channel} &   \textcolor{blue}{84.58±0.09} &  \textcolor{blue}{86.45 ±0.06} &  \textcolor{blue}{74.77±0.15} &  \textcolor{blue}{90.16±0.14} &  \textcolor{blue}{82.16±0.09}	 &  7.66±5.56 \\
   {multi-channel} &  56.84±0.22 & 66.51±0.14 & 43.27±0.23 & 49.69 ±0.29 & 81.02±0.11 & 8.23±5.85 \\
   \hline
   TSGM (Ours) & \\
   \hline
   {Flair(only)} &   58.58±0.24 &  64.23±0.19 &  45.33±0.24 &  53.72±0.29 &  73.92±0.13 &  \textcolor{blue}{6.37±4.42} \\		
   {T2(only)} & 67.88±0.17 & 74.37±0.14 & 53.48±0.17 & 89.24±0.19 & 57.19±0.20 & 18.11±8.92 \\
   {single-channel} &  \textcolor{red}{85.90±0.12} &  \textcolor{red}{86.84±0.09} &  \textcolor{red}{75.97±0.11} &  \textcolor{red}{91.02±0.09} &  \textcolor{red}{82.44±0.06} &  \textcolor{red}{5.08±3.26} \\		
   {multi-channel} & 81.76±0.14 & 84.22±0.10 & 71.08±0.16 & 86.24±0.14 & 81.60±0.09 & 7.46±5.15\\
   \bottomrule
\end{tabular}
}
\label{tab:ablation}
\end{table}

\subsection{Ablation Study}

We conducted two ablation studies to access the effectiveness of the proposed method. 

\begin{figure}
	\centering
	\includegraphics[width=\linewidth]{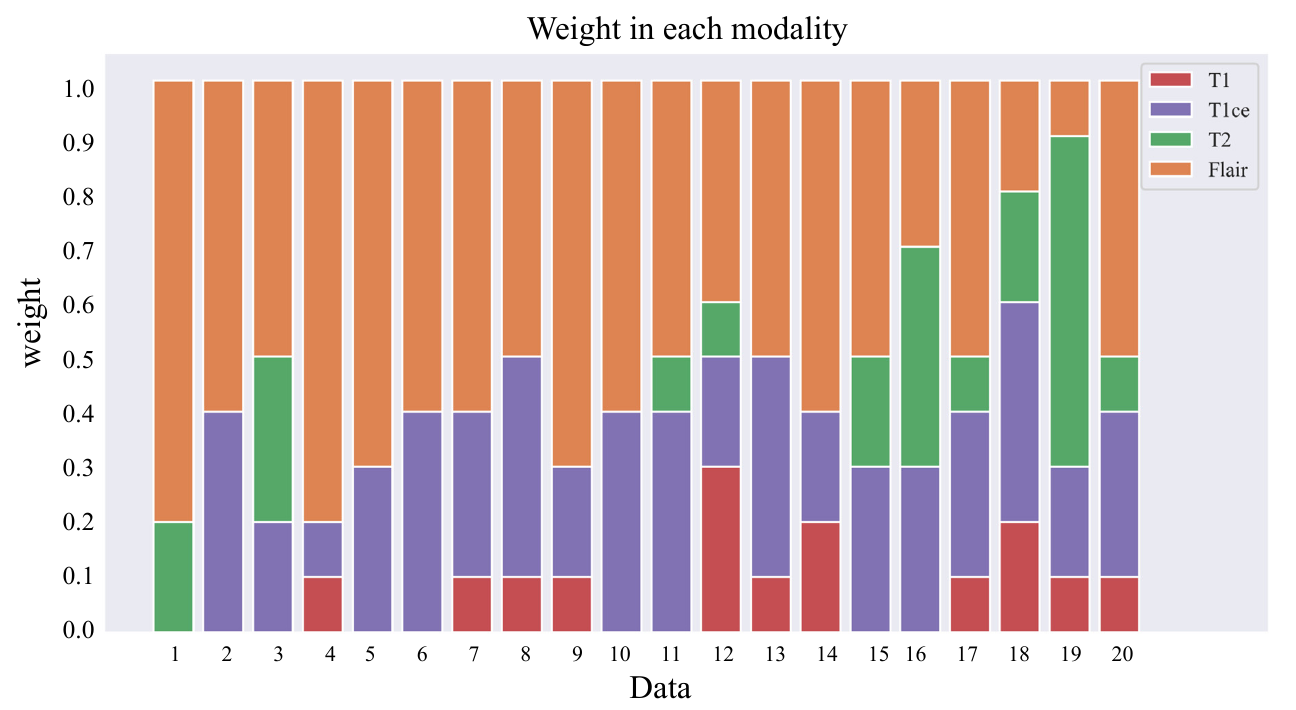}
	\caption{Illustration of the overlay histogram of each weight of T1w, T1ce, T2w, Flair under 20 sample data.}
	\label{fig:fig7}
\end{figure}

\textbf{Effect of different modalities and channels Input.} Clinical consensus suggests that Flair and T2 modalities are well suited for assessing the entire tumor region. During training, we considered three input configurations: Flair (only), T2 (only), and multimodal data. For the multimodal data, we experimented with a single-channel input combining multimodal data from different individuals (forming a matrix of size $1\times256\times256$) and a four-channel input consisting of 4 modalities data from the same individual (forming a matrix of size $4\times256\times256$ matrix).

\begin{figure*}
    \centering
    \includegraphics[width=0.7\textwidth]{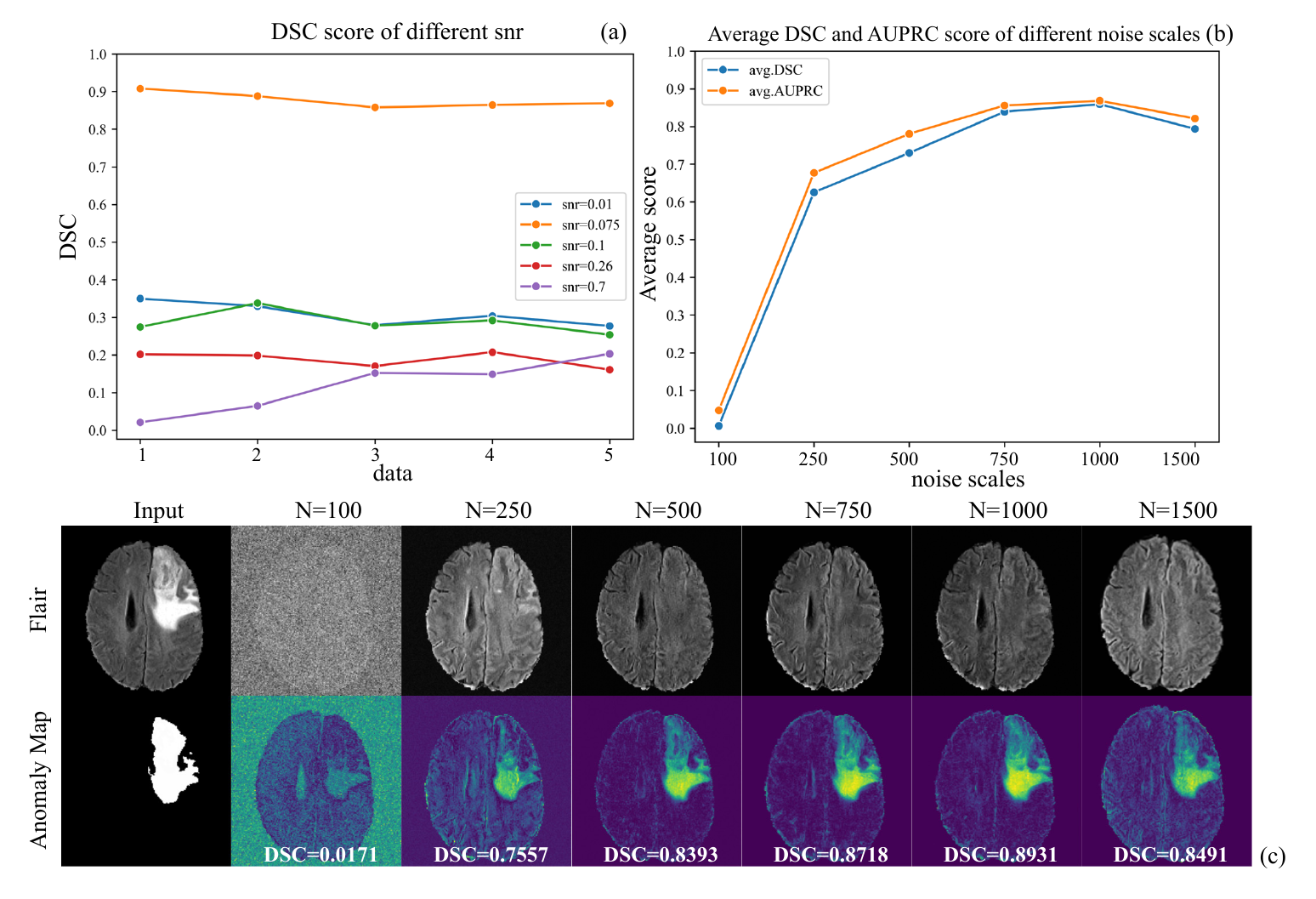}
    \caption{Change in segmentation quality as we vary scale of the snr or noise scale.}
    \label{fig:new1500}
\end{figure*}


When using multimodal data, the appropriate weights for [T1w, T1ce, T2w, Flair] in the training set were determined as described in section 2. Figure~\ref{fig:fig7} shows the overlay histogram of the final weights of each modality for 20 data points. It is observed that assigning higher weights to Flair, T2, and T1ce modalities, with T1w used as an additional complementary ensemble, resulted in more accurate detection of abnormal regions. The final proportion of each modality was calculated by averaging over all optimal weights as follows: Flair accounted for $44\%$, T2w for $30\%$, T1ce for $19\%$, and T1w for $7\%$.

The quantitative results of single-channel or multi-channel training are shown in table \ref{tab:ablation}. The results obtained by the four modalities ensemble outperformed those obtained by using only Flair or T2. Furthermore, the results of single-channel training under the two models are better than multi-channel.

\begin{figure*}
    \centering
    \includegraphics[width=0.7\textwidth]{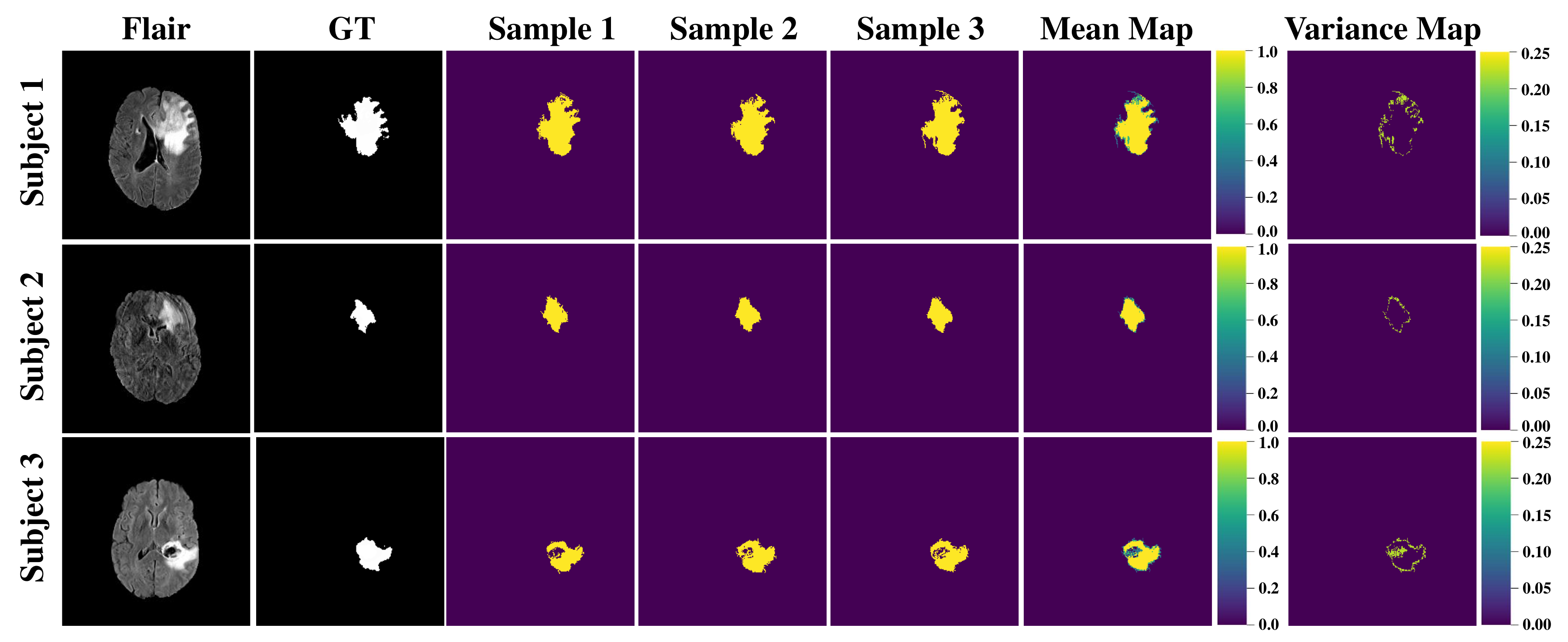}
    \caption{Mean and variance maps from 3 independent sampling runs.}
    \label{fig:v2}
\end{figure*}

\textbf{Effect of Hyperparameter.} Two hyperparameters control the strength of the guide in VE-JP: (1) The "signal-to-noise ratio" (snr) determined the step size $\epsilon$ in PC-sampling to control the balance of artifacts and noise. (2) the noise scale (N) governed the trade-off between model flexibility and sample fidelity. To investigate their impact on performance of the proposed method, we conducted a comparative analysis while varying the snr and N values. As depicted in Figure~\ref{fig:new1500}(a), the DSC values obtained from five representative data points vary as the snr changes, and we obtained the optimal snr of 0.075. Regarding the noise scale, Figure~\ref{fig:new1500}(b) and (c) present the corresponding average DSC and AUPRC scores, along with the reconstructed images obtained under different values of N. We observed that if N is too large, it may leads to a destruction of the images. If N is chosen too small, the model does not have enough freedom to remove the tumor from the image, and finally we get the optimal N is 1000.

\section{Discussion}

In this paper, we proposed the TSGM method for MRI-based brain anomaly detection. Our approach introduced a new conditional guidance to explore the potential of repairing the abnormal data and does not require labeled images or additional gradient trainers corresponding to a classifier for training. By synergistically fusing CycleGAN's ability to tackle well-posed healthy to diseased images translation, we can generate healthy-diseased paired images. By learning the joint distribution of healthy-diseased paired images using diffusion models as a prior, we design an iterative regularization approach to address the ill-posed translation from diseased to healthy images. Unlike CycleGAN, which directly learns the ill-posed mapping, our regularization method provides a more precise and stable solution for diseased to healthy image translation. The integration of CycleGAN and diffusion models empowers comprehensive and effective solutions for the intricacies of tumor detection and segmentation, overcoming key challenges like irregular lesion shapes and ambiguous boundaries. The experimental results indicated that TSGM performs better on conditional healthy image synthesis than other gradient-based guidance and outperforms alternative generative methods. That is because given the prowess of diffusion models in robust distribution learning, our conditional generation framework effectively captures the broader coverage of multi-modal healthy brain images distribution. For multi-modality images, it is verified that the segmentation performance of combined four modalities images with different weights outperforms that of any individual modality or any average four combined modalities. We also found that incorporating multi-channel modules did not yield significant improvements in segmentation performance. This could be attributed to the introduction of extraneous information from the multi-modalities, which potentially interfered with the final results. To address concerns over the non-deterministic sampling, we conducted three separate sampling runs and computed both the mean and variance maps of anomaly detection results, as shown in Figure~\ref{fig:v2}. The uncertainty analysis affirms that the randomness in our proposed diffusion model does not impede its effectiveness for diseased image restoration. Furthermore, we conducted ablation studies to demonstrate the optimality of each hyperparameter within the network architecture.

\begin{figure}
    \centering
    \includegraphics[width=1.\linewidth]{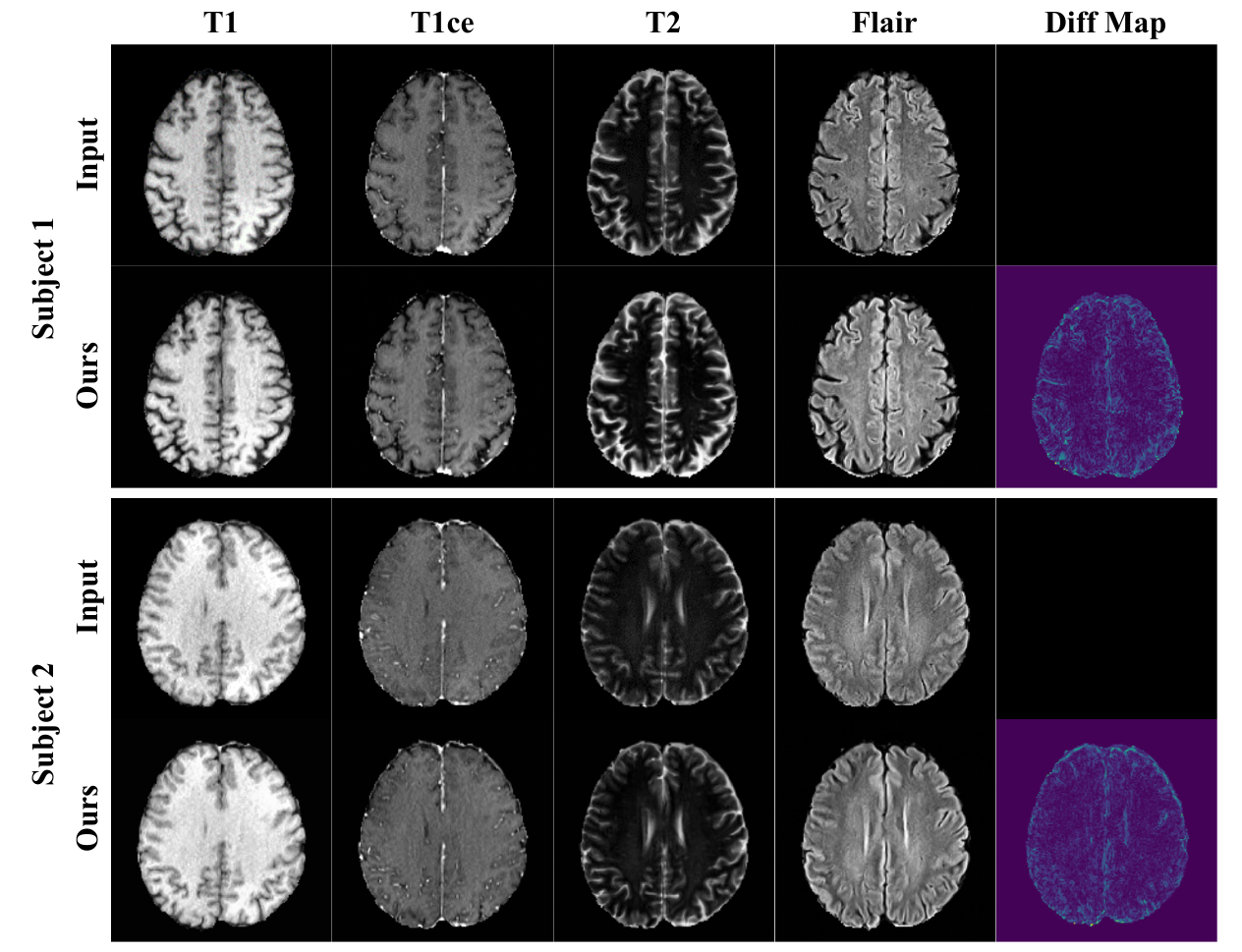}
    \caption{Results of the reconstruction in each modality for healthy subjects. First and third row: Original image. Second and fourth row: Reconstruction image and diff map.}
    \label{fig:health}
\end{figure}


In this study, we used different datasets to verify the proposed method. Although the training was merely conducted on the BraTs dataset, the method exhibits promising performance when directly applied to other datasets during testing. These results suggest that our approach demonstrates strong generalization capabilities and can be applied to segment other types of tumors.

For tumor detection, it is crucial to ensure that our method should not detect anomalies when applied to input images of healthy subjects. In Figure~\ref{fig:health}, we assess the effectiveness of our approach using two healthy slices from the BraTs dataset. Our findings demonstrate a highly intricate reconstruction of the image, yielding an anomaly map that approaches zero.

There are several limitations of this study. First, we noticed that the subtraction-based map is susceptible to interference from brainstem edges, potentially arising from inadequate edge detail representation during the reconstruction process. Future research can incorporate edge guidance to achieve a more realistic reconstruction of non-abnormal images. Second, we performed ensemble processing of multi-modality information after the models, future studies could explore the design of modules before inputting the models, such as downsampling to extract effective information for training. At the same time, it is worth mentioning that the current sampling speed is relatively slow, with an average restoration time of approximately 3 minutes per image. Hence, future work should focus on optimizing the sampling process to enhance efficiency, as well as harnessing the potential of multi-channel data more effectively.

\section{Conclusion}

We proposed a new framework named TSGM for brain tumor detection and segmentation. It achieves competitive performance in detecting and segmenting tumors on brain MR images in different datasets compared to conditional diffusion models and other generative models, and has the potential for brain tumor analysis in clinical application.  

\section*{Data availability statement}
This in-house dataset was currently the private property of Peking University Shenzhen Hospital. However, the deidentified images would be available upon reasonable request to corresponding author.

\section*{Acknowledgement}
This study was supported in part by the National Key R\&D
Program of China nos. 2020YFA0712200, National Natural Science Foundation of China under grant nos. 12226008, 81971611, Shenzhen Science and Technology Program under grant no.RCYX20210609104444089, JCYJ20220818101205012.

\bibliographystyle{IEEEtran}
\bibliography{refer}

\end{document}